\documentclass[a4paper]{jpconf}
\usepackage{graphicx, float, amsmath}
\graphicspath{{figures/}}
\begin{document}
\title{Pulse interactions in weakly nonlinear coherent optical communication links}

\author{J Gibney$^1$, I A Kuk$^2$ and I R Gabitov$^{1,2}$}

\address{$^1$ Department of Mathematics, University of Arizona, 617 N. Santa Rita Ave., Tucson, AZ 85721-0089, US}
\address{$^2$ Skolkovo Institute of Science and Technology, Bolshoy Boulevard 30, bld. 1, Moscow 121205, Russia}

\ead{ilya.kuk@skoltech.ru}

\begin{abstract}
The intrachannel interaction of pulses in weakly nonlinear coherent optical fiber lines is theoretically investigated. It is shown that the main contribution to the perturbation of the optical field comes from resonant interactions of ordered triplets of pulses. The structure of triplets is determined. The weight contributions of such interactions are calculated. A classification of interactions using L\"oschian numbers is proposed. Using  computer simulation, the dependence of the average energy of the optical field perturbations  on the distance is shown. Based on the performed analysis, an effective algorithm is proposed for assessing the perturbations  resulting from intrachannel interaction.
\end{abstract}

\section{Introduction}

The rapid development of the global information infrastructure is accompanied by a growing demand for bandwidth and spectral efficiency of optical communications. Currently, only optical fiber-based technology is able to keep up with the demands of high-speed telecommunications. Coherent fiber-optic data transmission is the chief technology in the field of high-speed communication, which came about due to significant progress in ultra-fast electronics, the development of the technology of optical system components and  implementation   of new   modulation formats of optical signals~\cite{leven2007coherent,kikuchi2015fundamentals}.

One of the basic principles underlying coherent communication is the assumption of the linear evolution of optical pulses as they propagate along an optical transmission line due to operating in a regime where nonlinear effects can be ignored.
The linear nature of coherent transmission is a factor which  strongly limits  further progress to increase the  data rate and the  length of transmission. Kerr nonlinearity is one of the natural parameters of  optical fiber, which results in nonlinear phase  modulation of the optical signal. This parameter introduces a characteristic length: nonlinearity length. Nonlinear distortion of the phase is of order one ($\varphi_{\text{nl}}\sim O(1)$)  after propagating this distance. Information encoding  in modern coherent communication systems is based on a differential phase modulation format. Nonlinear phase distortions lead to decoding errors and an increasing bit-error-rate ($BER$).   These distortions become noticeable when the length of the transmission line is comparable to the characteristic nonlinearity length.

Increase of the bit-rate ($BR$) also leads to nonlinear phase distortions. In order to retain an appropriate signal-to-noise ratio level, the optical pulse energy $\mathcal{E}$ must be greater than the threshold value $\mathcal{E} \geq \mathcal{E}_{\text{cr}}$. The pulse energy is estimated as $\mathcal{E} \sim \tau \times P$, where $\tau$ is the pulse width and $P$ is the characteristic power. This means that the pulse power increases with the bit rate as the bit rate is inversely proportional to the pulse width $P \geq \mathcal{E}_{\text{cr}}/\tau \sim \mathcal{E}_{\text{cr}} \times BR$.  Moreover, since the nonlinearity length $L_{\text{nl}}\sim1/(\alpha P)$ is inversely proportional to the product of the pulse power and the Kerr nonlinearity coefficient of the fiber $\alpha$, an increase in the system performance is  accompanied with  a decrease in the nonlinearity length. Demands for increasing  capacity  of communication systems for  rapidly developing information  technologies necessitates the principles of coherent communication to be extended to the nonlinear case. Among them is better understanding of the basics of nonlinear distortion of information carriers in optical fiber coherent systems. In this paper we consider the effects of nonlinear interchannel interactions on sequences of bits in coherent optical systems. We  consider the case of weak nonlinearity for moderate values of transmission length at which the nonlinear effects become noticeable.
\section{Principle of coherent communication }

\begin{figure}[h]
	\includegraphics[width=20pc]{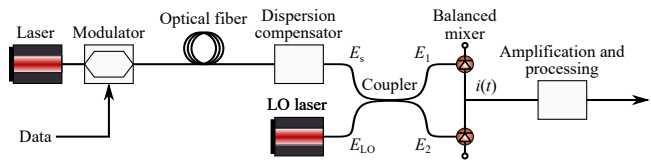}\hspace{1pc}
	\begin{minipage}[b]{16pc}\caption{\label{fig:coherent:scheme}Simplified schematic of a coherent communication line, here $E_{\text{s}}$ and $E_{\text{LO}}$ are optical fields of the signal and local laser oscillator, respectively}
	\end{minipage}
\end{figure}

A simplified schematic  of a coherent communication line is shown in Fig. \ref{fig:coherent:scheme}. Continuous laser output is converted by a modulator into a  train of pulses  with differential phase modulation. To illustrate the principle of differential phase keying, we can restrict ourselves to the simplest  case where the pulse phases have values of $0$ or $\pi$.  Decoding in the general case is carried out by comparing the phases of adjacent pulses (direct detection) or by comparing the pulse phase with the phase of continuous radiation of the local laser oscillator (coherent detection).
Dispersion broadening is compensated at the end of the transmission line, and in the case of coherent detection the signal $E_{\text{s}}$ is mixed with the continuous radiation of the laser oscillator $E_{\text{LO}}$  as shown in Fig.~\ref{fig:coherent:scheme}. After processing by a balanced mixer, an  optical signal $E_{\text{s}}$ is converted into an electrical current. The output current $i(t)$ of the balanced mixer is
\begin{equation}
	i(t) \sim \sqrt{P_{\text{s}} P_{\text{LO}}} \cos{\left[(\omega_{\text{s}} - \omega_{\text{LO}})t + \phi_{\text{s}}(t) - \phi_{\text{LO}}(t) \right]},
\end{equation}
where $P_{\text{s}}$ and $P_{\text{LO}}$ are powers, while $\omega_{\text{s}}$ and $\omega_{\text{LO}}$ are angular frequencies of the signal and local oscillator respectively. The phase $\phi_{\text{s}}(t)$ is modulated with the signal and $\phi_{\text{LO}}(t)$ contains only phase noise of a local oscillator that changes over time~\cite{kikuchi2015fundamentals}.

\section{Modeling equations}
The process of signal propagation in an optical fiber is approximately, and with sufficient accuracy, described by the nonlinear Schr\"{o}dinger equation. In this paper, we  consider the averaged  equation
\begin{equation}
	iE_z + \frac{1}{2} E_{tt}+ \varepsilon |E|^2 E=0,
	\label{eq:NLS}
\end{equation}
where $E=E(t,z)$ is the dimensionless amplitude of the electric field, $t$ and $z$ are the dimensionless retarded  time and distance, and $\varepsilon$ is the dimensionless coefficient of nonlinearity ($\varepsilon\ll 1$).  This equation is obtained by averaging the original model described in~\cite{hasegawa1991guiding} over the distance between the inline optical amplifiers compensating  optical fiber losses.
 As a boundary condition, we chose a sequence of Gaussian pulses (return to zero format) with Differential Phase Shift Keying (DPSK):
\begin{eqnarray}\label{Pulse:train}
	&& E(t,0) = \sum\limits_{k=1}^{N} a_{k}\pi^{-1/4} \exp\left[-(t-kT)^2/2\right], \\
	&& a_k = 1 \quad \hbox{with probability} \quad p_{1}=1/2 \nonumber \\
	&& a_k = -1 \quad \hbox{with probability} \quad p_{2}=1/2. \nonumber
\end{eqnarray}
Here, the energy of each pulse is normalized to one, the index $k$ is the number of a bit slot, and $N$ is the length of a sequence. Information in such a signal is encoded in the phase difference of adjacent pulses which is taken by the signs of coefficients  $a_k$. Since the problem contains a small parameter $\varepsilon$, we apply a perturbation method to the equation~(\ref{eq:NLS}). The  solution in this case has the following form:
\begin{equation}
	E(t,z)=E_0(t,z)+e(t,z).
\end{equation}
Linear evolution derived from equation~(\ref{eq:NLS}) with $\varepsilon = 0 $ is given by
\begin{eqnarray}
	E_0(t,z)= { \frac{1 }{\pi^{1/4} \sqrt{1+ i z}}}\sum_{k=1}^{N} a_{k}\exp\left[ -\frac{ (t-kT)^{2}}{2 (1+i z)}\right].
\end{eqnarray}
The equation for the perturbation, $e(t,z)$, is as follows:
\begin{eqnarray}
	i e_{z}+\frac{1}{2}e_{tt}=-\varepsilon |E_0 |^{2}E_{0},  \quad e(t,0)=0.
	\label{eq:perturb}
\end{eqnarray}

Solution of equation~(\ref{eq:perturb}) and the subsequent dispersion compensation  procedure (see Fig.~\ref{fig:coherent:scheme}) gives the following expression for the perturbation of the optical field in the $m^{th}$ slot (see also ~\cite{Gabitov2020}):
\begin{eqnarray} \label{Ghost:pulse}
	e_{m}(t,z)=i\varepsilon\pi^{-3/4} \sum_{k=1}^{N}\sum_{n=1}^{N} \int _{0}^{z}dx\frac{a_n a_k a_{k+n-m}^{*}}{ \sqrt{(x-i)(3x+i)}}\times \nonumber\\
	\exp \Biggl[-\frac{3 \left(x^2+1\right) }{2 \left(9 x^2+1\right)}\left(t-\frac{T \left(2 k+2 n+ m \left(3 x^2-1\right)\right)}{3 \left(x^2+1\right)}\right)^2 -\\ -\frac{T^2 \left(k^2+n^2 +m^2-k m- k n-m n\right)}{3 \left(x^2+1\right)}\Biggl]\times \nonumber\\
	\exp \Biggl[-i \frac{4 x}{9 x^2+1} \left(t-\frac{T}{4}  (3 k-2 m+3 n)\right)^2+ i \frac{T^2  (k-n)^{2}x}{4 \left(x^2+1\right)} \Biggl].\nonumber
\end{eqnarray}
This perturbation of an optical field is the result of nonlinear intrachannel interactions of overlapped pulses, which occurs due to dispersive broadening~\cite{Mamyshev:99}. Solution~(\ref{Ghost:pulse}) shows that with the growth of transmission distance this optical field is centred in the middle of the  time slot and the main contribution to this perturbation is provided by triplets of  optical pulses corresponding to time slots:   $k$, $n$, and $k+n-m$. In the case of dispersion management, this type triplet interaction was shown in~\cite{Ablowitz:00}. We assume that $a^{*}_{k+n-m}=0$ when $k+n-m>N$ or $k+n-m<1$ as the total number of slots is equal to $N$ and in these cases the triplets are positioned outside of the bit sequence and thus have to be neglected. The weight of triplet contributions is determined to be
\begin{equation}
	w_{k,n,k+n-m}(x)= \frac{e^{- \frac{T^2 \left(k^2 +n^2 +m^2 - km-kn-mn \right)}{3(1 + x^{2})} }}{\sqrt{(x^2 +1)(9 x^2+1)}}. \label{eq:weight}
\end{equation}
Since $k,~n,~m$ are integers, the quadratic form in (\ref{eq:weight}) is equal to L\"oschian numbers~\cite{Marshall75} $\lambda_j=0, 1, 3, 4, 7, 9, 12, 13, 16,\dots$:
\begin{equation}\label{Loesh:number}
k^2 +n^2 +m^2 - km-kn-mn=\lambda_j.
\end{equation}
Therefore L\"oschian numbers are ordering  groups of triplets according to their contribution to intrachannel interactions for every particular time slot.

\section{Intrachannel interactions}

\begin{figure}[h]
	\begin{minipage}{16pc}
		\includegraphics[width=16pc]{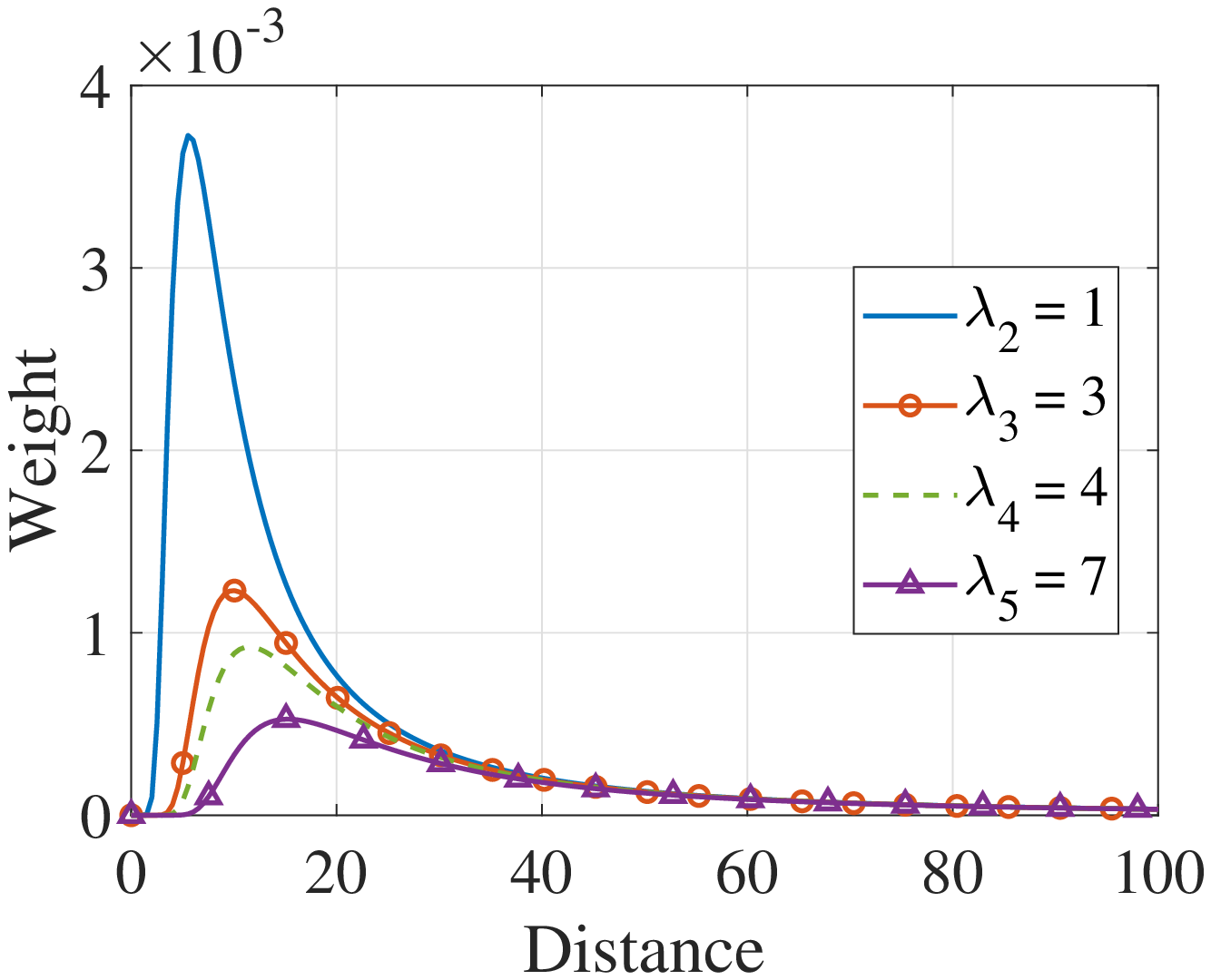}
		\caption{\label{fig:weight_vs_dist}Dependence of the weight factor on distance of propagation. Each curve corresponds to a certain L\"oschian number $\lambda_j$, which determines the weight.}
	\end{minipage}\hspace{2pc}
	\begin{minipage}{16pc}
		\includegraphics[width=16pc]{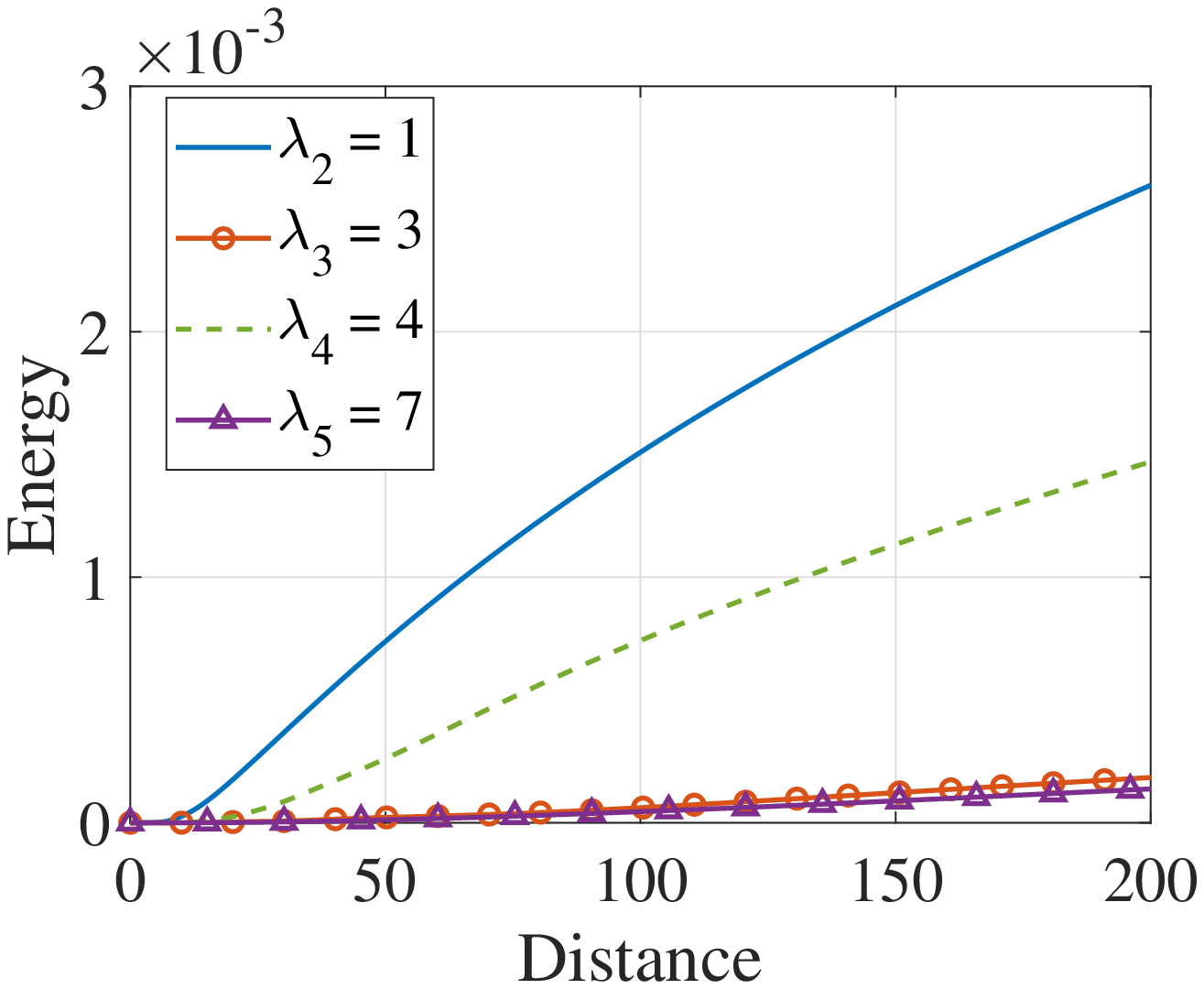}
		\caption{\label{fig:eng_vs_dist_100realiz}Here, the energy of nonlinear perturbations is calculated, taking into account only the contribution from certain sets of triplets, which are determined by the L\"oschian number $\lambda_j$.}
	\end{minipage}
\end{figure}

From the formula~(\ref{Ghost:pulse}) it follows that the result of intrachannel interactions in each slot is determined by the resonant interaction of a certain set of triplets of pulses. The contribution of each triplet is determined by the corresponding L\"oschian number and to each L\"oschian number corresponds  a set of triplets with the same weight factor. Further, for convenience, the order of numbering in the sequence of pulses has been changed -- the number of the examined slot $m$ is set to zero, $m = 0$. Dependence of the weight factors of triplets corresponding to  different  L\"oschian numbers is shown in Fig.~\ref{fig:weight_vs_dist}. The   pulse energies at $z=0$ in (\ref{Pulse:train})  are normalized to one and  the size of the bit slot is, for these simulations, chosen to be  $T=10$.

 Figure~\ref{fig:weight_vs_dist} shows that at  $x=0$, when pulses are not overlapping, there is no intrachannel interaction and weight factors are zero. At the initial stage of propagation along the fiber, dispersive broadening leads to an increase of overlapping, resulting in the growing strength of nonlinear intrachannel interaction.  After reaching its maximal value, each  weight factor is monotonically decaying since amplitudes of the pulses are decreasing with broadening and thus the nonlinear interaction is decreasing as well. Figure~\ref{fig:weight_vs_dist} also shows that  higher L\"oschian numbers have lower  weight, since they correspond to pulse triplets farther from the slot under consideration, and therefore will be of lower magnitude by the time of overlap. This also shifts the maximum to a larger distance. The highest weights are corresponding to $\lambda = 1$. This means that interactions between slots ``$0$" and  ``$\pm 1$" and contribution to the ``$0$" slot from  interaction between slots ``$-1$" and ``$-2$" as well as from interaction between slots  ``$1$" and ``$2$" are strongest and have the same weight.

The energy of nonlinear perturbations in slot $m$ is determined as
\begin{equation}
	\mathcal{E}_m(z) = \int_{T(m-1/2)}^{T(m+1/2)} dt |e_{m}(t,z)|^{2}
\end{equation}
and strongly depends on the structure of the bit-pattern which is encoded in terms of pulse phases (signs of coefficients $a_{k}$). To analyze the average contribution to the total energy by the triplets corresponding to different L\"oschian numbers, we modified the expression (\ref{Ghost:pulse}) and performed summation over only the triplets with  L\"oschian numbers under consideration and averaged the result over a randomly generated ensemble of bit-pattern realizations.

Figure~\ref{fig:eng_vs_dist_100realiz} shows the dependance of the average energy of nonlinear perturbations caused  only from a particular set of triplets specified  by the L\"oschian numbers. Here, averaging over 100 realizations of random 128-bit sequences has been performed for a slot located in the middle of the pulse train. The dimensionless nonlinearity coefficient has been chosen to be $\varepsilon=0.01$.  Figure~\ref{fig:eng_vs_dist_100realiz} shows that, on average, the main contributions come from the triplets corresponding to L\"oschian numbers $1$ and $4$. The first set of  resonant  triplets ($\lambda =1$) contain near neighbor slots and four of them include the slot under investigation. The second  set of  resonant  triplets ($\lambda =4$) also contains  four triplets  that include a slot under investigation. The relative contribution of other slots is much smaller (the difference is an order of magnitude). It should be emphasized that the contribution of the resonant triplets corresponding to the value of $\lambda = 3$ is much less than the contribution of the resonance triplets   corresponding to $\lambda = 4$, despite the fact that the components of the resonant triplets of the first group are located closer to the slot under consideration. This is due to the fact that the triplets of the first group do not contain the analyzed slot itself.

The set of triplet groups characterized by L\"oschian numbers contains two subgroups. The first subgroup contains triplets corresponding to interactions of the pulse located in the analyzed slot (in our notations $0^{th}$ slot) with one of the pulses within the train in the $k^{th}$ slot ($k\neq 0$). In this case, both  $m=0$ and $n=0$, therefore $\lambda =k^2$ in accordance to~(\ref{Loesh:number}). Hence if $\lambda$ is a L\"oschian number which  is equal to the square of an integer, $\lambda =k^2$, then the corresponding triplets are indicating the presence of interacting  ``$0$" and  ``$\pm k$" slots and contribution to the ``$0$" slot from the  interaction of ``$\pm k$" and ``$\pm 2 k$" slots. The second subgroup contains triplets which provide only contributions to the ``$0$" slot and do not interact directly with the pulse located in the ``$0$" slot.

Figure~\ref{fig:eng_vs_dist_100realiz} shows that the growth of the average energy perturbation field in the  ``$0^{th}$" slot is primarily determined by the interactions of the first subgroup. To illustrate this, we performed numerical experiments using four different approaches labeled below as Algorithms 1-4. In each case, a bit-pattern is randomly generated and the perturbation energy in the ``$0^{th}$" slot is computed by evaluating the summation (\ref{Ghost:pulse}) over a subset of triplet interactions.

\begin{figure}[h]
	\begin{minipage}{16pc}
		\includegraphics[width=16pc]{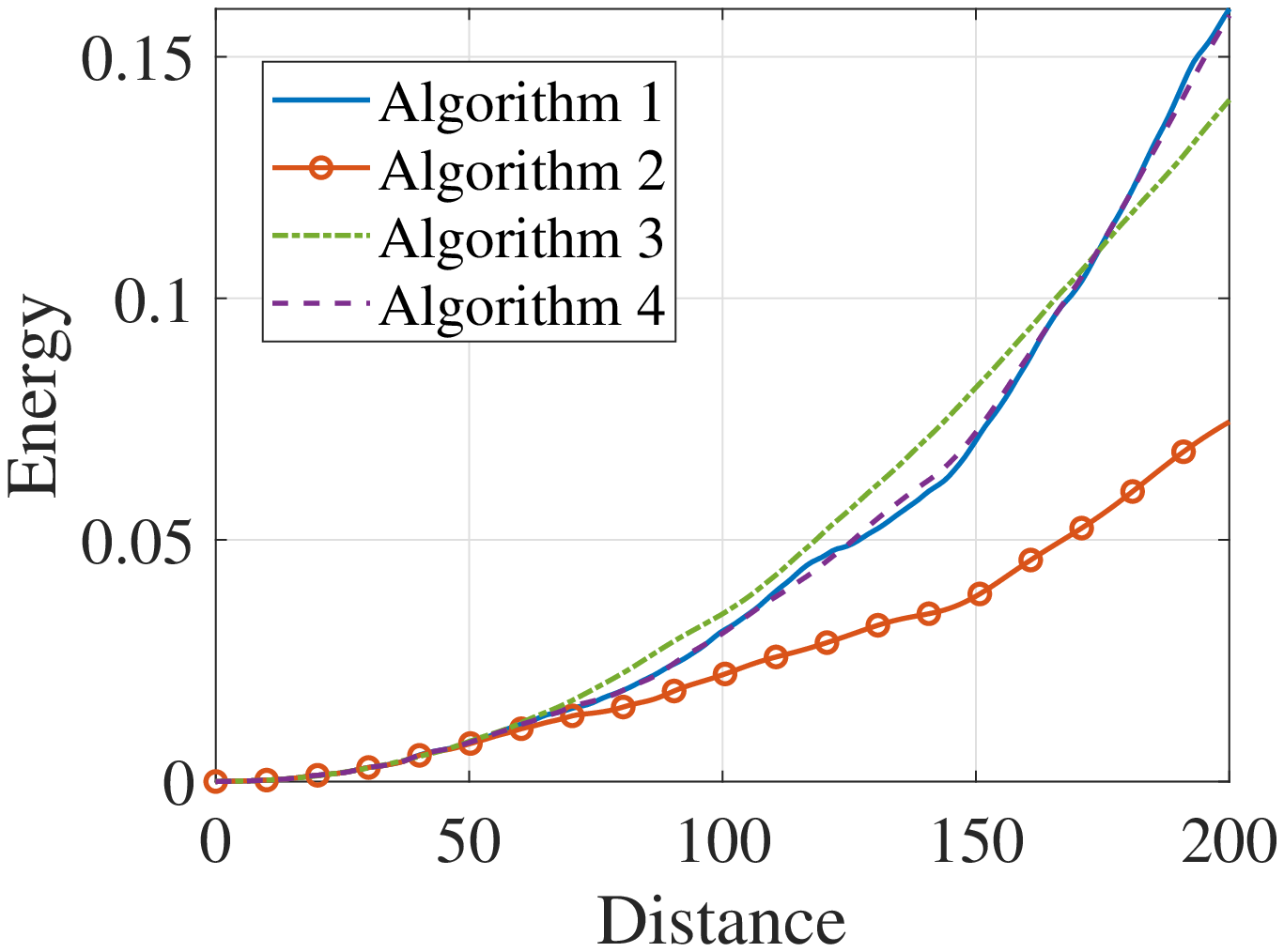}
		\caption{\label{fig:algorithms}Comparison of four different  algorithms used to evaluate nonlinear perturbations.}
	\end{minipage}\hspace{2pc}
	\begin{minipage}{16pc}
		\includegraphics[width=16pc]{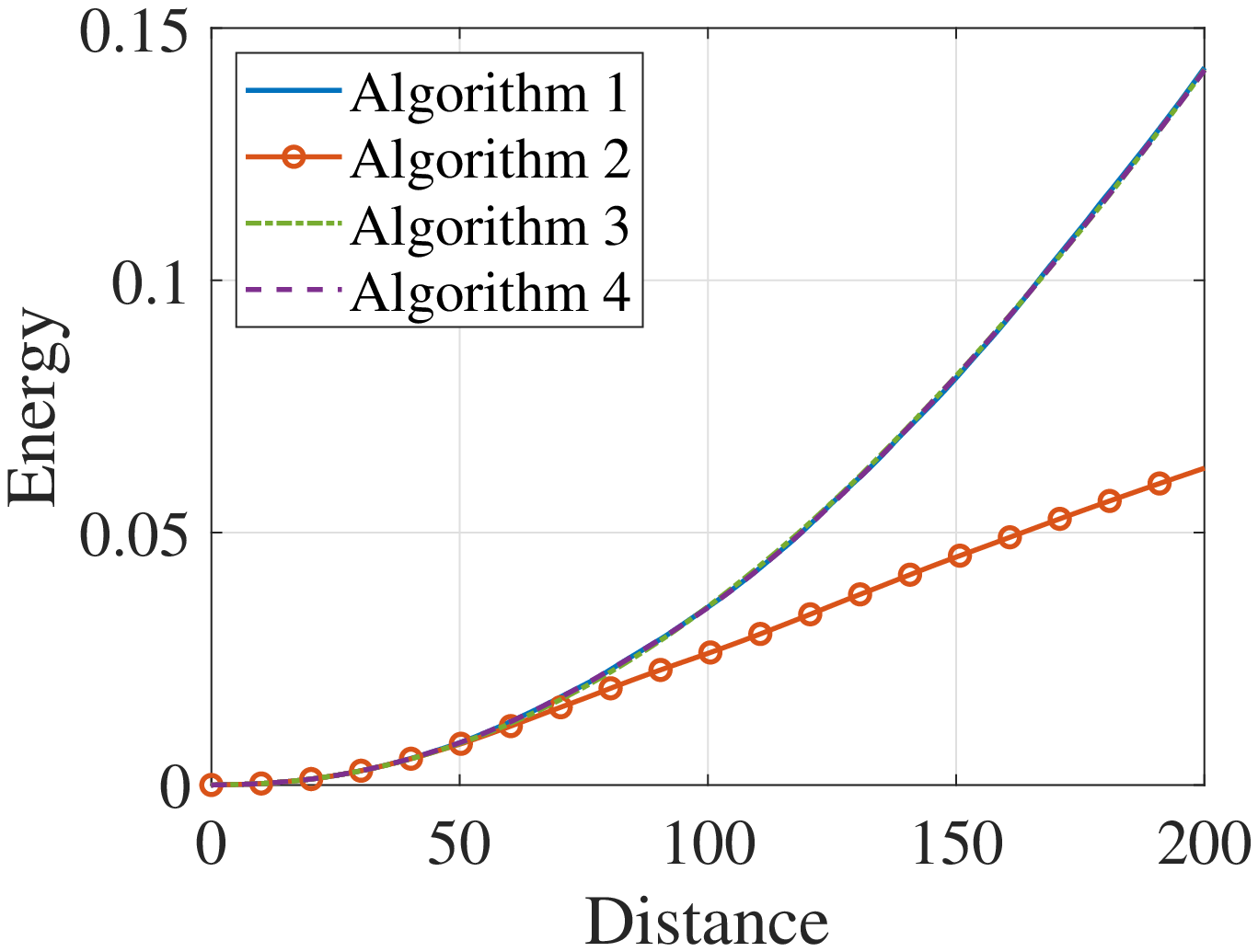}
		\caption{\label{fig:algorithms:100reliz}Monte-Carlo simulation averaging over 100 random realizations of a bit sequences. Curves corresponding to  algorithms 1, 3 and 4 are practically  the same.}
	\end{minipage}
\end{figure}

\textbf{Algorithm 1:}
In the first case, summation of (\ref{Ghost:pulse}) is performed over all $k$ and $n$, therefore  all possible interactions are considered. The result of this simulation is shown in Fig.~\ref{fig:algorithms} as the solid blue line labeled as ``Algorithm~1".  

\textbf{Algorithm 2:}
In the second case, we consider only triplets corresponding to the set of L\"oschian numbers $\lambda_{j} $, $j = 1,2, \dots, M,$ where we have chosen $M = 31$ as the cutoff. This corresponds to considering all interactions, both direct and indirect, but only up to a distance of approximately 9 slots. The result of this simulation is shown in Fig.~\ref{fig:algorithms} as the solid orange line with open circle labeled as ``Algorithm~2". 

\textbf{Algorithm 3:}
In the third case, the density of L\"oschian numbers was reduced and only a set of L\"oschian numbers representable as a square of natural numbers was used for summation. This corresponds to considering only the direct interactions discussed previously. The corresponding curve is labeled as ``Algorithm~3" and is shown by dash-dotted green line. 

\textbf{Algorithm 4:}
In the fourth case, we use a hybridization of Algorithms 2 and 3 -- resonant triplets corresponding to all L\"oschian numbers up to $\lambda_{31}$ are considered, and then above that only L\"oschian numbers representable as squares of a natural number. This corresponds to considering all interactions up up a distance of approximately 9 slots, and then only direct interactions at larger distances. Results of simulations are presented by a purple dashed line and labeled as ``Algorithm~4".

Figure~\ref{fig:algorithms} shows very good agreement between all approaches in the interval $0\le z \le 60$. For $z$ values greater than 60 ($z\ge 60$), Algorithm 4 stands out as the most accurate, with the best match to the benchmark Algorithm 1. In this case, the whole set of resonant interactions between a group of pulses in the interval between approximately -9 and 9 slots corresponds to the full set of  L\"oschian numbers up to $\lambda_{31}=79$. The characteristic pulse width at a distance of $z \sim 60$ is approximately $\tau \sim 90$. This means that when $ z \sim  60 $, the broadened pulse covers approximately $ \sim 4-5 $ slots on each side. Considering that the peripheral slots are also broadened  and cover the same number of slots on each side, we come to the conclusion that approximately $9-10$ slots are involved in the interaction on each side of the zero slot, which is consistent with the value of L\"oschian number $\lambda_{31}=79$. In this case, approximately $10$ slots are involved in direct interaction with the zero slot. With an increase in the distance $ z $ and the same restriction on the L\"oschian number, the number of directly interacting slots increases to $ \sim 20 $, and then the growth of the energy of the perturbation field increases only due to an increase in indirect interactions which have low weight. Hybridization of Algorithm 2 and Algorithm 3, realized  in Algorithm 4, shows that taking into account only direct interactions outside   $ \lambda_ {31} $ makes it possible to model intrachannel interactions with good accuracy while reducing the computation time by a factor of approximately 7. Table~\ref{tab:time} illustrates the results of a comparative analysis of the computation time in each numerical experiment.

In order to verify that it is sufficient to take indirect interactions into account only within $\sim 8-9 $ of the nearest slots on each side, we analyzed the growth of the averaged energy of the perturbation field as a function of the distance $z$. The result is shown in Figure~\ref{fig:algorithms:100reliz}. Averaging was performed over an ensemble of 100 random realizations   of bit patterns. The result of calculations shows a practical coincidence of the curves corresponding to Algorithms 1, 3 and 4.

\begin{table}
	\caption{\label{tab:time}Comparison of execution time of algorithms}
	\begin{center}
		\begin{tabular}{lllll}
			\br
			Algorithm & 1 & 2 & 3 & 4\\
			\mr
			Time& 100\% & 7.4\% & 7.7\% & 14.6\%\\
			\br
		\end{tabular}
	\end{center}
\end{table}

\section{Conclusion}
The intrachannel interaction of pulses in a quasilinear coherent communication line is theoretically studied. The resonant interaction of ordered pulse triplets  is shown to make the main contribution to the perturbation of the optical field. The weights of the contributions of interactions of this type are determined. Proposed  classification of such interactions is  based on the use of L\"oschian number. It is shown that direct type interactions are sufficient to approximate the perturbations of a sequence of pulses arising from intrachannel interaction.

\section*{References}
\bibliography{iopart-num}
\end{document}